\documentclass[a4paper,fleqn,usenatbib]{mnras}

\usepackage{newtxtext,newtxmath}

\usepackage[T1]{fontenc}
\usepackage{ae,aecompl}

\usepackage{graphicx}	
\usepackage{amsmath}	
\usepackage{amssymb}	

\usepackage{hyperref}
\hypersetup{
  citecolor=cyan,      
}

\usepackage{graphicx}
\usepackage{color}
\definecolor{orange}{rgb}{1,0.5,0}

\title[High-order MHD in magnetised binary neutron stars]{Beyond
  second-order convergence in simulations of magnetised binary neutron
  stars with realistic microphysics}

\author[E.R. Most et al.]{Elias R. Most$^{1,2}$\thanks{Corresponding author: emost@itp.uni-frankfurt.de},
L. Jens Papenfort$^{1}$, L. Rezzolla$^{1,3}$
\\
$^{1}$ Institut f\"ur Theoretische Physik, Goethe Universit\"at Frankfurt am Main, Germany\\
$^{2}$ Center for Computational Astrophysics, Flatiron Institute, 162 Fifth Avenue, New York, NY 10010, USA \\
$^{3}$ School of Mathematics, Trinity College, Dublin 2, Ireland}

\date{Accepted XXX. Received YYY; in original form ZZZ}

\pubyear{2019}

\begin{document}
\label{firstpage}
\pagerange{\pageref{firstpage}--\pageref{lastpage}}
\maketitle

\begin{abstract}
We investigate the impact of using high-order numerical methods to study
the merger of magnetised neutron stars with finite-temperature
microphysics and neutrino cooling in full general relativity. By
implementing a fourth-order accurate conservative finite-difference
scheme we model the inspiral together with the early post-merger and
highlight the differences to traditional second-order approaches at the
various stages of the simulation. We find that even for
finite-temperature equations of state, convergence orders higher than
second order can be achieved in the inspiral and post-merger for the
gravitational-wave phase. We further demonstrate that the second-order
scheme overestimates the amount of proton-rich shock-heated ejecta, which
can have an impact on the modelling of the dynamical part of the kilonova
emission. Finally, we show that already at low resolution the growth rate
of the magnetic energy is consistently resolved by using a fourth-order
scheme.
\end{abstract}


\section{Introduction}

In the era of gravitational wave and multi-messenger astronomy of binary
neutron stars accurate numerical modelling of neutron-star mergers and
their remnants on long timescales $\simeq 1\, \rm s$ has never been more
important. The coincident detection of a gravitational-wave signal from a
neutron-star merger \citep{Abbott2017} and an accompanying
electromagnetic counterpart in form of a short gamma ray burst
\citep{Abbott2017d} and kilonova afterglow \citep{Kasen2017,Drout2017}
has established a firm connection with electromagnetic counterparts and
highlights the need for multiphysics modelling of neutron-star mergers. A
neutron-star merger consists of several stages starting from the late
inspiral, where accurate numerical waveforms are needed to calibrate
analytical models for waveforms, through merger
\citep{Kawaguchi2018,Nagar2018,Dietrich2019}, which requires
sophisticated microphysics in terms of finite temperature equations of
state (EOS) satisfying recent observational constraints
\citep{Annala2017,Most2018,Tews2018,Burgio2018,Raithel2018} until the
post-merger phase where neutrino and magnetic viscosity can drive large
amounts of mass ejection
\citep{Just2015,Siegel2017,Siegel2018,Fernandez2018, Fujibayashi2017b},
that are needed to make a connection to the kilonova afterglow produced
by the decay of heavy elements in the matter outflow. The modelling of
this complicated multi-physics system requires both the use of numerical
relativity \citep{Baiotti2016, Duez2019} and of an accurate modelling of
the fluid, the electromagnetic fields as well as the
microphysics. Considerable effort has been placed on improved and highly
accurate methods to solve Einstein field equations numerically
\citep{Baumgarte2010,Shibata_book:2016} and to couple them with
high-order methods for relativistic hydrodynamics \citep{Radice2013c,
  Bernuzzi2016}. At the same time, mainly driven by an effort to model
core-collapse supernovae, very sophisticated numerical schemes for
neutrino transport have been developed
\citep{Ruffert96b,Buras06b,Shibata2011,Sumiyoshi:12,Foucart2015a,
  Just2015}. When considering the late stages of the evolution of the
system not only is it important to account for the various relevant
physics contributions, such as neutrino interactions, but it is also
crucial to understand how numerical errors at finite numerical resolution
accumulate over time. This is even highlighted by the fact that current
simulations of neutron-star mergers sometimes show non-convergent
behaviour in the magnetic field evolution \citep{Endrizzi2016,Ciolfi2017}
even when small resolution changes are used. Notwithstanding, that with
current computational efficiencies and available resources not even all
relevant physical scales involving magnetic turbulence can be resolved
\citep{Kiuchi2015a, Kiuchi2017}, studying the late time evolution of the
remnants accretion disk is not only feasible but has been the subject of
recent investigations \citep{Siegel2017,Siegel2018,Fernandez2018}. While
all such simulations so far have used traditional second-order accurate
finite volume schemes to model the evolution of the general-relativistic
magnetohydrodynamics system (GRMHD), earlier works have already indicated
the benefit of using more accurate high-order methods in this context
\citep{DelZanna2007,tchekhovskoy_2007_wham,Radice2013c}, while even more
recent studies have already started to consider advanced finite-element
approaches \citep{Kidder2016,Fambri2018}. Taking an intermediate approach
similar to \citep{DelZanna2007,mccorquodale2011high,Chen2016,Felker2018}
we will consider the impact of using a fourth-order accurate numerical
scheme to model the merger of magnetised binary neutron stars and show
the advantages gained when additionally finite-temperature effects and
neutrino cooling are included.

\section{Formulation and Methods}

In this study we solve the GRMHD equations in dynamical spacetimes (see
\citealt{Duez05MHD0,Giacomazzo:2007ti} for an overview), together with
realistic microphysics. We will start by giving a brief overview of the
system of equations and will then discuss details of the implementation.

The space-time is described by the metric\\ ${\rm d}s^2 = g_{\mu\nu} {\rm
  d}x^\mu{\rm d}x^\nu$ decomposed using the 3+1 split of spacetime
\citep{Alcubierre:2008, Rezzolla_book:2013}, that can be written as
\begin{align}
ds^2 = \left( -\alpha^{2} + \beta_i \beta^i \right) dt^2 +
2 \beta_i dx^i dt +
\gamma_{jk}dx^j dx^k\,.
\end{align}

The fluid is described by the baryonic rest-mass density $\rho$, the
electron fraction $Y_e$ and the temperature $T$. Using these quantities
the pressure $p\left( \rho,T,Y_e \right)$ and the specific internal
energy $\epsilon\left( \rho,T,Y_e \right)$ can be computed from the
equation of state (EOS), which closes the set of equations. Although this
study highlights the benefits of high-order schemes for neutron-star
simulations with realistic finite-temperature EOS, the latter equally
applies also to approaches using simpler EOS.

The dynamics of the fluid is governed by the fluid four velocity $u_\mu =
\left( u_0, W v_i \right)$, $v_i$ is the three velocity of the fluid and
$W = \sqrt{1- v_i v^i}^{-1}$ the Lorentz factor.

The evolution of the electromagnetic field is described by the Faraday
tensor $F_{\mu\nu} = \nabla_\mu A_\nu - \nabla_\nu A_\mu$, where $A_\mu =
\left( \Phi, A_i \right)$ is the vector potential, and using the Lorenz
gauge \citep{Etienne2012a} $\nabla_\mu A^\mu=0$ the Maxwell equations in
the ideal MHD limit can be written as \citep{Baumgarte2003, Etienne2012a}
\begin{align}
  \partial_t A_i - \epsilon_{ijk}\tilde{v}^j B^k + \partial_i\left(
  \alpha \Phi -\beta^j A_j \right) =& 0\,, \\ \partial_t \left(
  \sqrt{\gamma} \Phi \right) + \partial_j\left( \alpha\sqrt{\gamma} A^j
  -\beta^j \sqrt{\gamma} \Phi \right) =& - \xi \alpha \sqrt{\gamma} \Phi\,,
  \label{eqn:Maxwell}
\end{align}
where $\sqrt{\gamma} B^i = \epsilon^{ijk}\partial_j A_k$ is the magnetic
field as seen by the normal observer, $\epsilon^{ijk}$ is the totally
antisymmetric Levi-Civita tensor, $\tilde{v}^i =\alpha v^i - \beta^i$ is
the transport velocity and $\xi^{-1}$ is the damping time scale of the
gauge variable. It is also useful to define the magnetic field in the
fluid rest-frame, i.e., $4\pi\, b_\mu = \left( g_{\mu\nu} +
u_{\mu}u_{\nu} \right) B^\nu$.

The dynamics of the fluid is governed by the energy momentum tensor
$T_{\mu\nu}$ \citep{Baumgarte2003,Shibata05b}
\begin{align}
  T_{\mu\nu} = \left(\rho\left( 1 +
  \epsilon \right) + b^2 + p\right)u_\mu u_\nu + \left( p +
  \frac{b^2}{2} \right)g_{\mu\nu} - b_\mu b_\nu\,,
  \label{eqn:Tmunu}
\end{align}
which gives rise to the equations of hydrodynamics via $\nabla_\mu
T^{\mu\nu}= 0$. Using the 3+1 split these equations can be recast as
follows \citep{Baumgarte2003,Duez05MHD0,Shibata05b,Giacomazzo:2007ti}
\begin{align}
  \partial_t \rho_\ast + \partial_j \left( \rho_\ast \tilde{v}^j \right)
  =& 0 \label{eqn:GRMHD0} \\ \partial_t S_i + \partial_j \left( \alpha
  \sqrt{\gamma}\ T^{j}_i \right) =& \frac{1}{2}\alpha
  \sqrt{\gamma}\ T^{\mu\nu} \partial_i g_{\mu\nu} \\ \partial_t \tau +
  \partial_j \left( \alpha^2 \sqrt{\gamma}\ T^{0j} - \rho_\ast
  \tilde{v}^j \right) =& \alpha \sqrt{\gamma} \left[\left( T^{00} \beta^i
  \beta^j \right. \right. \nonumber\\ & \left. \left. + 2 T^{0i} \beta^j
  + T^{ij} \right)K_{ij} \right. \nonumber\\ & \left. - \left(
  T^{00}\beta^i + T^{0i} \right)\partial_i \alpha \right]\,,
  \label{eqn:GRMHD}
\end{align}
which are already written in conservative form using
\begin{align}
  & \rho_\ast = \sqrt{\gamma} \rho W\,, \label{eqn:rhostar}\\
  & S_i =  \left( \rho_\ast hW + \sqrt{\gamma} B^2 \right) - \left( B^j v_j
  \right) \sqrt{\gamma}B_i\,, \\
  & \tau = \rho_\ast \left(hW -1\right) -
  \sqrt{\gamma} p + \sqrt{\gamma}B^{2} - \frac{\sqrt{\gamma}}{2}\left[
  B^2 \left( 1-v^2 \right) + \left( B^i v_i \right)^2
  \right]\,,
  \label{eqn:tau}
\end{align}
given as function of the primitive quantities $\rho,v^i,\epsilon$.
Here $h = 1 + \epsilon + p/\rho$ is the specific enthalpy.

\subsection{Numerical methods}

In the following we will briefly present the numerical methods used in
this work. We will start by reviewing how to obtain high-order schemes
for the GRMHD equation, before discussing the various parts necessary for
numerically stable high-order GRMHD simulation involving realistic EOS
and neutrino cooling.

\subsubsection{High-order HRSC methods}

The GRMHD equations need to be solved using high resolution shock
capturing (HRSC) methods in order to address discontinuities that may
appear both in the magnetic field as well as in the hydrodynamic
variables, see \cite{Toro09} for an overview. Such equations are solved
in discretized forms, i.e., an equation of the form
\begin{align}
  \partial_t U + \partial_x F =0\,,
  \label{eqn:U}
\end{align}
with state $U$ and flux $F$ is recast into
\begin{align}
  \partial_t \hat{U} + \frac{1}{\Delta x} \left[ \hat{F}_{i+1/2} -
    \hat{F}_{i+1/2} \right] =0\,,
  \label{eqn:Udisc}
\end{align}
where $\hat{U}$ and $\hat{F}$ are suitable discretisations on a
computational grid $x^i$. Typically such equations are solved in a
finite volume form, where
\begin{align}
  \hat{U}= \int_\mathrm{cell} \mathrm{d}V\ U ~ ~ ~ ~ ~ \hat{F} =
  \int_\mathrm{face} \mathrm{d}A\ F\,
  \label{eqn:FV}
\end{align}
are evaluated over a cell centered on $x^i$ and its adjacent faces,
respectively. At second order the integral can be approximated by the
value at the midpoints, i.e., $\hat{U}^i = U^i$ is evaluated at the
centre $x^i$ and $\hat{F}^{i +1/2}$ is computed at the interface between
$x^i$ and $x^{i+1}$. The values of $U$ necessary to compute $F$ are then
computed by simply reconstructing $U$ to the cell interface using a
limited interpolation $\mathcal{R}$ that is shock-aware, and then
applying a suitable Riemann solver to provide suitable upwinding in the
computation of $\hat{F} = F\left( \mathcal{R}\left[ U \right]
\right)$. This is the most common approach and is implemented in most
simulation codes for compact systems
\citep{Etienne2015,Duez05MHD0,Giacomazzo:2007ti,Foucart2013a,Liebling2010,Moesta13_GRHydro}.
A different approach consists in interpreting \eqref{eqn:Udisc} in a
finite-difference sense, i.e.,
\begin{align}
\hat{U}= U\,, ~ ~ ~ ~ ~ \hat{F} = \mathcal{R}\left[ F \right]\,,
  \label{eqn:FD}
\end{align}
where $\mathcal{R}$ is a limited interpolation, similar to the
reconstruction of $\hat{U}$ in the finite volume approach, to which an
additional upwinding formula similar to a Riemann solver in the previous
case is applied \citep{Shu2003,mignone_2010_hoc}. The main difference
between the two cases is that the latter case is a direct update of point
values using point values, where the order of the method is solely set by
the order of the interpolation routine $\mathcal{R}$, whereas in the
finite volume sense the order is additionally set by the numerical
approach used to evaluate the volume and surface integrals in
\eqref{eqn:FV} \citep{mccorquodale2011high, Felker2018}.

Typically, such a finite difference approach requires a characteristic
decomposition of $U$ at every step in order to reduce spurious
oscillations \citep{Shu2003}. We have in fact found that when not
performing this step while using a realistic EOS, it can lead to large
oscillations and unphysical states in low density regions. Instead, we
opt for a different approach, the so-called ECHO scheme
\citep{DelZanna2007,DelZanna2003,DelZanna2002}, in which \eqref{eqn:FD}
is split in a reconstruction and (high-order) derivative step
$\mathcal{D}$, such that
\begin{align}
  \hat{F} = \mathcal{D}\left[ F\left( \mathcal{R}\left[ U \right] \right)
    \right]\,.
  \label{eqn:FD_echo}
\end{align}
Here, first the primitive variables are reconstructed to the cell
interfaces, then a suitable Riemann solver is used to compute the flux
$F$ and an additional correction step $\mathcal{D}$ is applied, such that
\eqref{eqn:Udisc} is a high-order approximation to \eqref{eqn:U}. Most
importantly, at second order this step can be omitted and the above
finite volume scheme is recovered. At higher orders this is not the case
and at fourth order one finds \citep{DelZanna2007}
\begin{align}
  \hat{F} = F\left( \mathcal{R}\left[ U \right] \right) - \frac{\Delta x^2}{24}
  F''\left( \mathcal{R}\left[ U \right] \right)\,,
  \label{eqn:FD_DZ}
\end{align}
where the second derivative can be approximated using the standard finite
difference expression
\begin{align}
  F''_{x_{i+1/2}}=\frac{F_{x_{i+3/2}} + F_{x_{i-1/2}} - 2
    F_{x_{i+1/2}}}{\Delta x^2}\,.
\end{align}

It should be remarked that this fixed stencil choice can in principle
induce oscillations near shocks, but in practice is extremely robust as
has been shown in \citep{DelZanna2007, Chen2016}. On the other hand such
a choice is necessary for constraint transport schemes as the flux
gradient and the curl operator used to compute the magnetic field from
the vector potential have to always cancel out to machine
precision. Although it is possible to also use the fourth derivative
$F^{\left( 4 \right)}$ to increase the formal convergence order to 6, we
will restrict ourselves in this work to fourth-order as it has been shown
in previous (formally) high-order purely hydrodynamical studies that e.g
convergence of the phase shift in gravitational waves converges at most
at third order \citep{Radice2013b,Radice2013c,Bernuzzi2016}.

\subsubsection{Implementation} \label{sec:fil}

We solve the coupled Einstein-GRMHD system using the
\texttt{Frankfurt/IllinoisGRMHD} code (\texttt{FIL}), which is a high
order extension of the publicly available \texttt{IllinoisGRMHD} code
\citep{Etienne2015} that is part of the \texttt{Einstein Toolkit}
\citep{loeffler_2011_et}. In the following we will give an overview of
the numerical and implementation details.

To solve the Einstein equations \texttt{FIL} provides its own spacetime
evolution module, which implements the Z4c
\citep{Hilditch2012,Bernuzzi:2009ex} and CCZ4 \citep{Alic:2011a,Alic2013}
formulations using forth order accurate finite differencing
\citep{Zlochower2005:fourth-order} with different choices for the
conformal factor. In this work we choose $\psi^{-2}$ and adopt the Z4c
formulation with a damping coefficient $\kappa=0.02$
\citep{Weyhausen:2011cg, Hilditch2012}. The space-time gauges are evolved
using the standard 1+log slicing and shifting-shift Gamma driver
conditions \citep{Alcubierre:2008, Baumgarte2010, Rezzolla_book:2013},
where a uniform damping parameter of $\eta = 2/M$ is adopted. The
implementation makes use of modern compile-time evaluated C++14 templates
provided by the \texttt{TensorTemplates} library and can be vectorised
using parallel datatypes \citep{Kretz2011}.

The GRMHD equations \eqref{eqn:GRMHD0} - \eqref{eqn:GRMHD} are solved
using the ECHO scheme \citep{DelZanna2007} as given in \eqref{eqn:FD_DZ},
making our code overall formally fourth-order accurate. The fluxes are
computed from the reconstructed primitive variables
$(\rho,\ T,\ Y_e,\ Wv^i,\ \sqrt{\gamma}B^i)$ using a HLLE Riemann solver
\citep{Harten83}. The reconstruction step $\mathcal{R}$ is performed
using the WENO-Z method \citep{Borges2008}, with the optimal weights and
stencils for a conservative finite difference scheme taken from
\citep{DelZanna2007}. We point out that these are different from those
for finite volume or traditional finite difference approaches
\citep{Borges2008}. The magnetic fields are evolved via a vector
potential $A_i$ and the gauge field $\Phi$. The use of a vector potential
trivially allows us to generalise the ECHO scheme to a multi-grid
context, since the magnetic field is recomputed at every iteration from
the vector potential $A_i$, thus preserving the divergence constraint. In
order to maintain high-order convergence in this step the derivative
correction $\mathcal{D}$ also needs to apply in this context,
specifically we compute
\begin{align}
  \sqrt{\gamma}B^x_{i+\frac{1}{2},i,k} =
  &-\frac{\mathcal{D}\left[A_{i+\frac{1}{2},j,k+\frac{1}{2}}\right] -
    \mathcal{D}\left[A_{i+\frac{1}{2},j,k-\frac{1}{2}}\right]}{\Delta z}
  \\ \phantom{=}&+\frac{\mathcal{D}\left[A_{i+\frac{1}{2},j+\frac{1}{2},k}\right] -
    \mathcal{D}\left[A_{i+\frac{1}{2},j-\frac{1}{2},k}\right]}{\Delta y}\,.
  \label{eqn:BfromA}
\end{align}
Different from \citep{Etienne2015} we implement the upwind constraint
transport scheme as in \citep{DelZanna2007} in which the staggered
magnetic fields $\sqrt{\gamma} B^i$ are reconstructed from two distinct
directions to the cell edges, which we found greatly minimises diffusion
related to a dimensional bias compared to the original implementation
\citep{DelZanna2003} when used with a high-order scheme. The
cell-centered magnetic fields are always interpolated from the staggered
ones using fourth-order unlimited interpolation in the i-th direction for
$B^i$, in which this magnetic field component is continuous
\citep{londrillo2004divergence}. A potential downside of high order
schemes is that they can cause oscillations at sharp discontinuities,
e.g., at the surface of the neutron star. In order to remedy such
behaviours we follow the approach of \citep{Radice2013c} and hybridise
the high-order flux with a first order local Lax-Friedrichs flux based on
a positivity preserving criterion for the conserved density $\rho_\ast$
\citep{Hu2013}. It should be remarked that this step has to be done
before the derivative correction \eqref{eqn:FD_DZ} is applied, in order
to be compatible with discrete $\nabla \cdot \mathbf{B}$-constraint
\citep{DelZanna2007}. While this no longer exactly guarantees positivity
of the conserved density, since the hybridisation coefficient is computed
without accounting for the $\mathcal{D}$ correction, we find that
stability is, nonetheless, improved and failures at the primitive
inversion stage are greatly reduced allowing for a stable evolution of
the neutron-star surface during the inspiral.

\subsection{Microphysics}

In light of the kilonova detection AT2017gfo \citep{Cowperthwaite2017}
modelling the mass ejection and its microphysical composition has become
crucial for any study of neutron-star mergers including at least a basic
treatment of weak interactions and finite-temperature EOS. Consequently,
the \texttt{FIL} code includes a framework to handle tabulated
finite-temperature EOS either in the
StellarCollapse\footnote{\url{https://stellarcollapse.org}} or in the
CompOSE format \footnote{\url{https://compose.obspm.fr}}. Any quantity
$Q\left(\rho, T, Y_e\right)$ in the EOS is then treated by means of a
three-dimensional linear interpolation. Equilibrium weak interactions are
computed following the approach laid out in
\citep{Ruffert96b,Rosswog:2003b} and implemented in
\citep{OConnor10,Galeazzi2013,Neilsen2014}, including
\begin{itemize}
  \item (inverse) $\beta$-decay
  \item electron-positron pair annihilation
  \item plasmon decay
  \item neutrino scattering on heavy nuclei and free nucleons
  \item electron-flavor neutrino absorption on free nucleons
\end{itemize}

Presently, neutrino interactions are included via a leakage approach
\citep{Rosswog:2003b, Sekiguchi2010CQG, Deaton2013, Galeazzi2013}, which
accounts for the effects of neutrino cooling by means of a neutrino
emissivity $Q_{\nu_i}$ and neutrino emission rates per baryon
$R_{\nu_i}$. Since these depend sensitively on the optical depth
$\tau_{\nu_i}$, we distinguish between optically thin (free-streaming)
and optically thick (diffusive) regimes by computing a harmonic average
of the rates following \citep{Rosswog:2003b}. The optical depth
$\tau_{\nu_i} = \int \kappa_{\nu_i}\,{\rm d}s$ is computed from the
opacities $\kappa_{\nu_i}$ by locally applying Fermat's principle to the
neighbouring cells on the computational grid
\citep{Neilsen2014,Foucart2013a}. The energy- and momentum-loss due to
neutrino emission is then included by modifying the evolution equations
accordingly \citep{Sekiguchi2010,Deaton2013, Galeazzi2013}
\begin{align}
  \partial_t \left( \rho_\ast Y_{e} \right) + \partial_i \left( \rho_\ast
  Y_e \tilde{v}^i \right) =&-\alpha \sqrt{\gamma} \rho \left( R_{{\nu}_e}
  - R_{\bar{\nu}_e} \right), \\ \partial_t S_i + \dots =& -\alpha
  m_b^{-1} \rho_\ast \mathcal{Q} v_i ,\\ \partial_t \tau
  +\dots=& -\alpha m_b^{-1} \rho_\ast \mathcal{Q}\,,
\end{align}
where $Q = \sum_i Q_{\nu_i}$ and $m_b$ is the baryon mass.

\subsection{Primitive inversion}
\label{app:inversion}

A particularly delicate part is the non-analytical computation of the
primitive variables from the evolved conserved variables
\eqref{eqn:rhostar} - \eqref{eqn:tau}. In the past a great effort has
been devoted to designing stable and accurate inversion algorithms
\citep{Noble2006,Mignone2007,DelZanna2007,Etienne2012,Muhlberger2014,Newman2014,Palenzuela2015,Nouri2018}.
Similar to a recent review \citep{Siegel2018a} of these inversion schemes
we perform a multi-stage inversion that consists of the following steps

\begin{enumerate}
  \item First check $ \varepsilon > B^2/\rho_{\ast}$, where $\varepsilon$
    is a very small number so that the contribution of $B^2$ to $\tau$
    and $S^2$ can be neglected to around machine precision if the above
    condition is fulfilled. In this case we are certainly in a purely
    hydrodynamical regime and can apply the algorithm of
    \citep{Galeazzi2013} for general relativistic hydrodynamics for which
    an exact bracketing for the 1D-root finding problem can be provided
    and solved efficiently using Brent's method \citep{Brent2002}. If we
    have not found a root within the given accuracy we proceed to the
    last step.
  \item If magnetic fields cannot be neglected we proceed with the single
    variable algorithm of \citep{Palenzuela2015}. Before performing the
    root-finding we apply the generalised solvability constraints
    \citep{Palenzuela2015,Etienne2012}. If the root-finding has converged
    we assess the quality of the solution by computing (A4) of
    \citep{Mignone2007} and check if the relative error on $\tau$ is at
    least $\sim 10^{-12}$. If the inversion was not successful or too
    inaccurate we proceed with the next step.
  \item If we were not successful in the previous step we perform the
    fixed-point inversion of \citep{Newman2014}. We have found that when
    disabling the Aitken acceleration step proposed in \citep{Newman2014}
    the robustness of the scheme increased in demanding situations. If a
    fixed-point has been found, perform the same consistency check on the
    solution as in the previous step.
  \item Finally, if all previous attempts were unsuccessful we replace
    the energy variable $\tau$ by the specific entropy $s$ per baryon
    that was independently evolved at every step as in \citep{Nouri2018}
    \begin{align}
	    \partial_t \left( \rho_\ast s \right) + \partial_i
            \left(\rho_\ast s \tilde{v}^i \right) = \frac{m_n \alpha
              \sqrt{\gamma}}{k_B T} \left[ Q_\nu - R_\nu \mu_{\nu_e}
              \right]\,,
	    \label{eqn:s_evol}
    \end{align}
where $m_n$ is the neutron mass and $\mu_{\nu_e}$ the electron neutrino
chemical potential. Following \citep{Nouri2018} we solve (A24) of
\citep{Muhlberger2014} using Brent's method. Note that for this step the
pressure $P = P(\rho)$ is effectively a function of the density alone
since for given $\rho$ and $Y_e$ in the iteration the temperature $T =
T\left( \rho, s, Y_e \right)$ can be trivially recovered and from it the
pressure $P= P \left( \rho,T,Y_e \right)$.
  \item If also the entropy cannot be used to recover a point, we reset
    it to atmosphere values. This happens only on very rare occasions
    close to atmosphere values.
\end{enumerate}

While the above procedure might seem in total rather involved and
potentially expensive we find that the use of a fully high-order flux
update greatly reduces the need for applying the entropy fix. This can be
understood by considering the core part of every inversion algorithm, the
computation of $\epsilon$ at every step, see e.g., \citep{Palenzuela2015}
for an expression. While for example the ideal fluid $p = \rho \epsilon
\left( \Gamma -1 \right)$ every positive value of $\epsilon$ corresponds
to a physical pressure, this is no longer the case when using a finite
range tabulated EOS, see Fig. 1 in \citep{Galeazzi2013}. Since at every
intermediate step the recovered value of $\epsilon \left( \rho,T,Y_e
\right)$ has to be inverted for the temperature $T$, using a wrong and in
most cases too low value for $\epsilon$ results in the recovery of a
wrong temperature $T$ and hence of a wrong enthalpy $h_\text{recovered} <
h$, which is smaller than the actual enthalpy $h$. We note that the same
is also true when using the evolved entropy $\tilde{s}$ as this always
corresponds to a lower temperature $T\left(\tilde{s},\rho_s,Y_e\right) <
T$, as its evolution equation \eqref{eqn:s_evol} is only correct in the
absence of shocks, and will hence always lead to smaller entropy values.

\begin{figure*}
  \centering \includegraphics[width=0.45\textwidth]{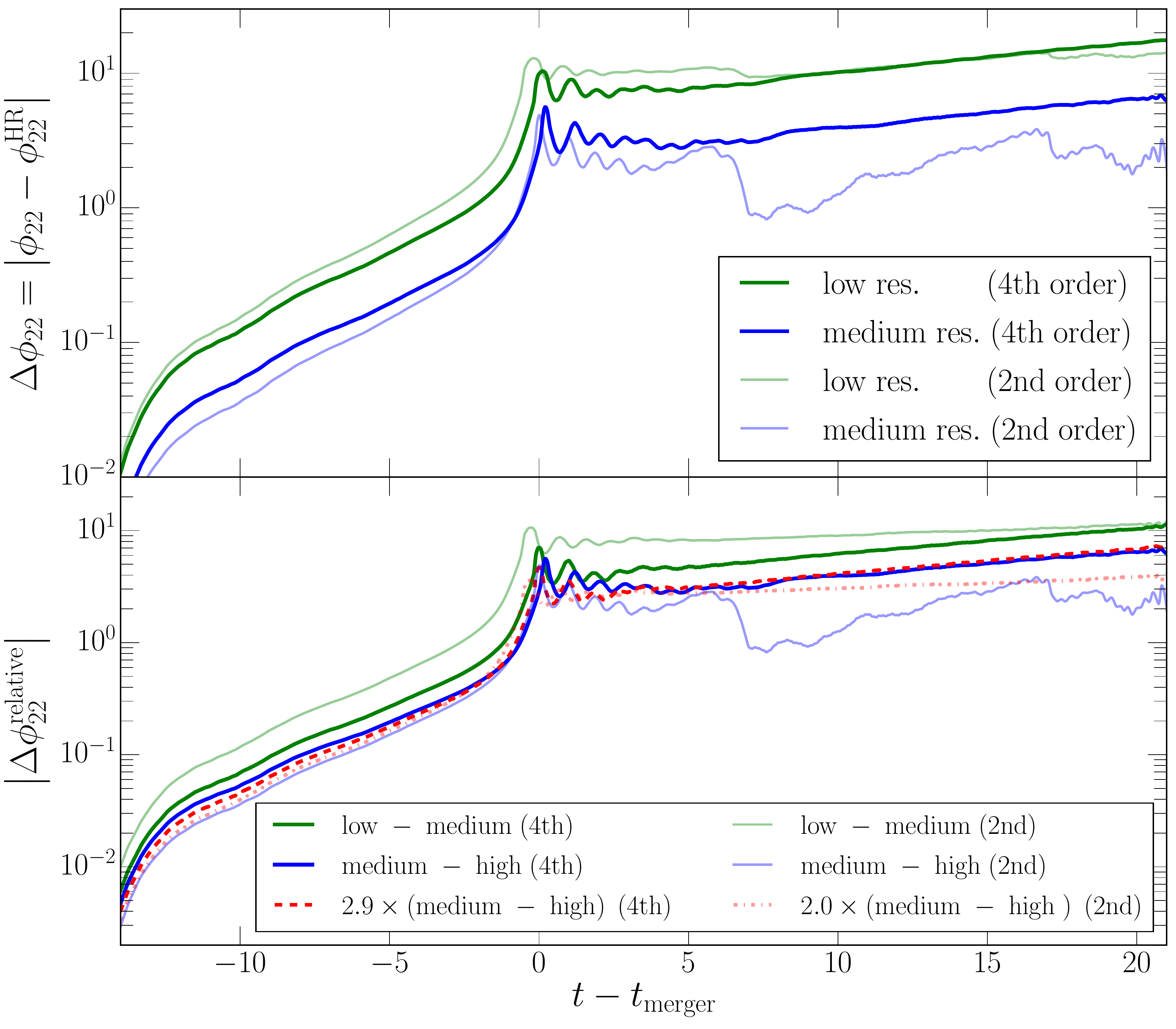}
  \hskip 0.5 cm
  \centering \includegraphics[width=0.45\textwidth]{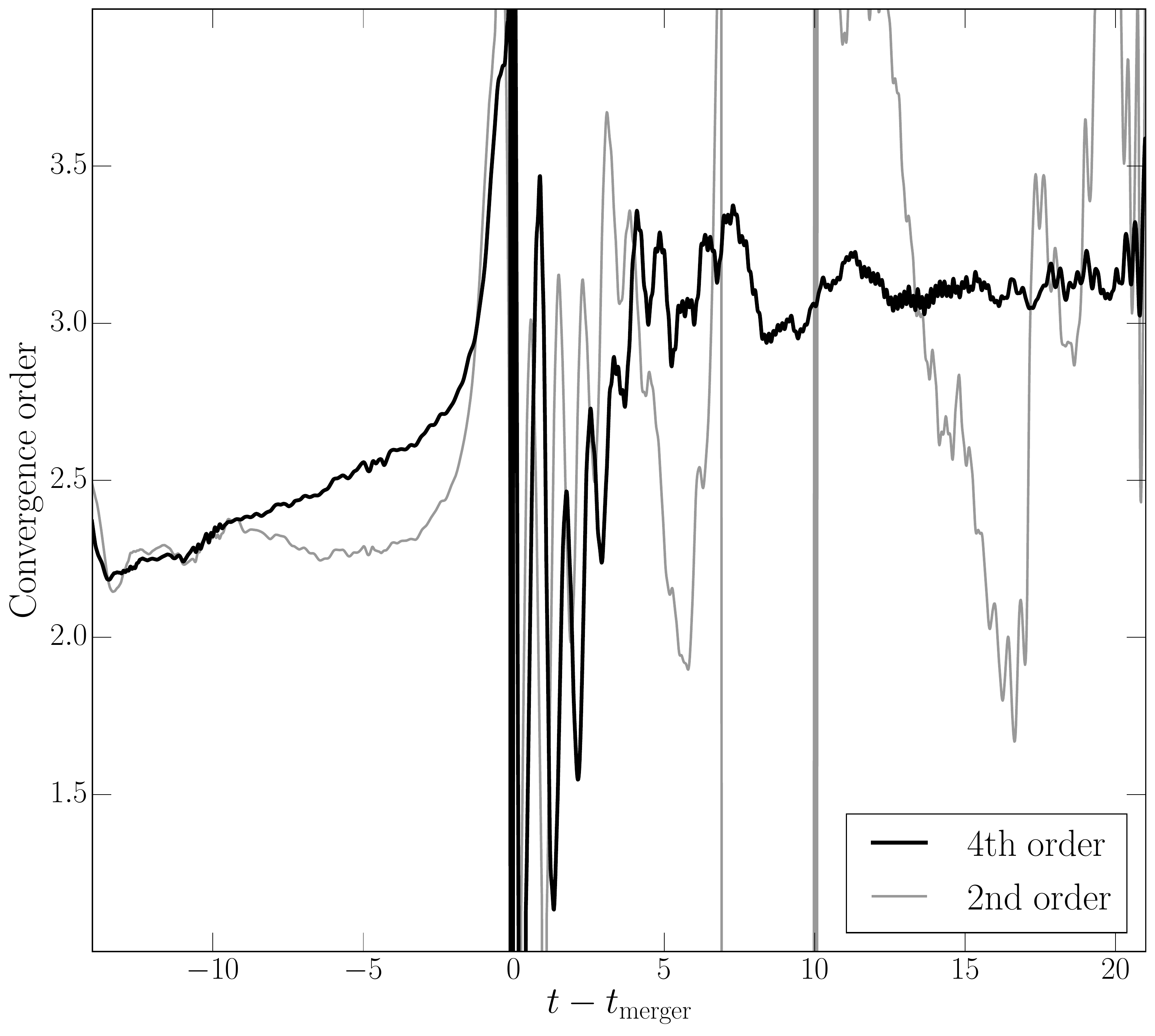}
    \caption{ {\it Left:} (Top) Phase difference $\Delta \phi_{22}$ of
      the $\ell=m=2$ mode to the high resolution model $\phi_{22}^{\rm
        HR}$ and relative differences $\Delta \phi_{22}^{\rm
        relative}$. The semi-transparent line denotes the second-order
      model. {\it Right:} Self convergence order of the phase $\phi_{22}$
      using the three available resolutions.}
  \label{fig:GW_TNTYST}
\end{figure*}

\begin{figure}
  \centering \includegraphics[width=0.49\textwidth]{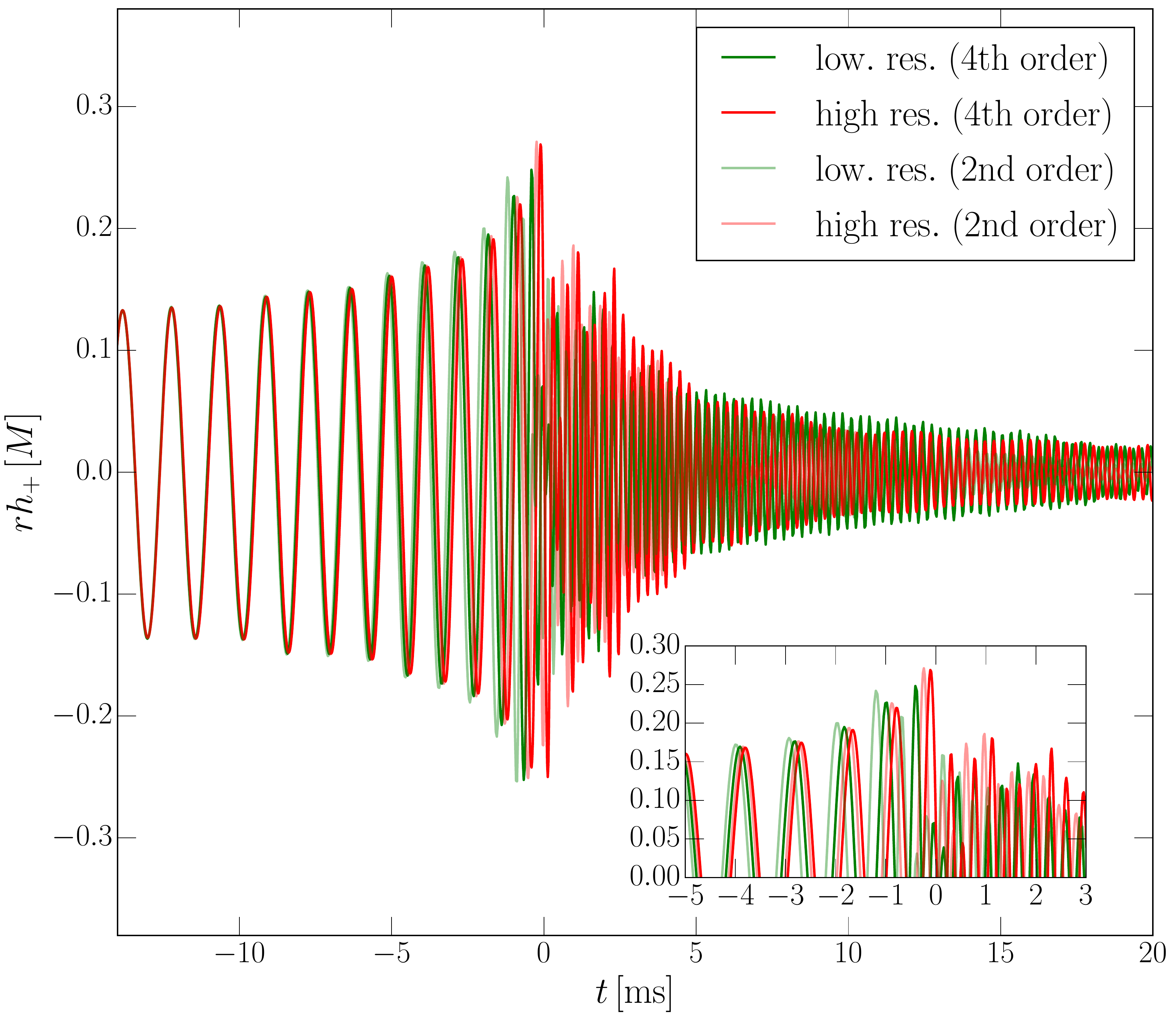}
  \vspace{15mm}
  \centering \includegraphics[width=0.49\textwidth]{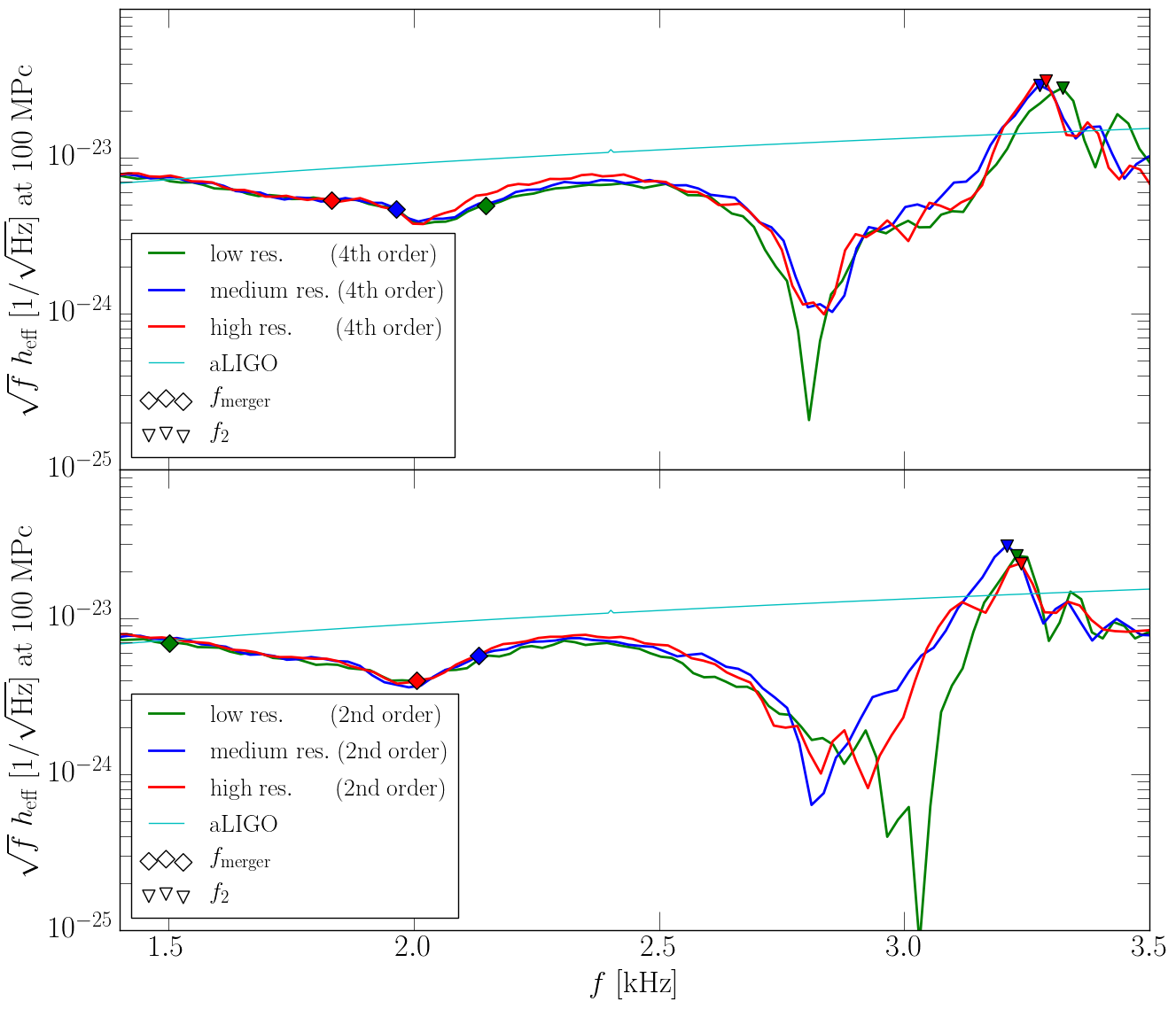}
    \caption{ {\it Top:} Gravitational-wave strain $(\ell=m=2)$ mode for
      a source at 100 Mpc extrapolated to spatial infinity for different
      resolutions ranging from $0.1328-0.25\, M_\odot$.  {\it Bottom:}
      Gravitational wave spectrum showing the effective strain
      $\tilde{h}_{\rm eff}\left( f \right)$ computed using the high-order
      scheme (top) and the second-order scheme (bottom). The cyan solid
      line represents the sensitivity curve of aLIGO.}
  \label{fig:GW_TNTYST_strain_heff}
\end{figure}

\section{Results}

In the main part of this work we focus on the dynamics of an irrotational
equal mass neutron-star binary with a total mass of $2.7\ M_\odot$
constructed using the \texttt{COCAL} code
\citep{Tsokaros2015,Tsokaros2018}. The two stars are initially placed at
a distance of $45\,\rm km$ and are equipped with a poloidal magnetic
field with a maximum field strength $B_c\simeq 5\times 10^{14}\,\rm G$
confined to the interior of the two stars. The simulation domain is
modelled by a series of seven nested boxes extending up to $\simeq
1500\,{\rm km}$. We consider a total of three resolutions $194\, \rm m$,
$262\,({\rm 2nd})/295\,({\rm 4th})\, \rm m$ and $370\, \rm m$, in the
following referred to as \textit{high (HR), medium (MR)} and \textit{low
  (LR)} resolutions. The composition of the stars is described by the
soft TNTYST EOS \citep{Togashi2017}, which is consistent with recent
constraints on the EOS from GW170817 \citep{Most2018}.

In the following we will describe the inspiral and post-merger dynamics
of the systems and highlight the differences between formally second and
fourth-order convergent numerical schemes. In particular, we will compare
two sets of simulations using exactly the same methods as described in
Sec. \ref{sec:fil}, but where for one set the high order correction
\eqref{eqn:FD_DZ} is switched off. This approach using high order
reconstruction combined with an HLL Riemann solver at overall
second-order convergence is commonly used when modelling of neutron-star
mergers and their remnants \citep{Reisswig2012b,Muhlberger2014,
  Foucart2015b,Bovard2017,Nouri2018, Radice2018a, Papenfort2018} and
hence allows us to draw conclusion directly relevant for existing and
future studies of neutron-star mergers.

The merger of a neutron-star binary can be separated into several stages,
including the inspiral, early post-merger and long-term post-merger
phase. Each stage gives rise to different observables and sets the
initial conditions for the next stage of the evolution. Being able to
accurately model each of these is a key task for numerical-relativity
codes and in the following we will look at the first to stages in light
of the impact of fully high-order numerical schemes.

\subsection{Accurate modelling of neutron-star inspirals}

An important aspect of numerical-relativity codes is their ability to
compute gravitational waveforms for compact binaries. Since analytic
waveform models rely on numerical-relativity input to account for tidal
deformations in the last orbits, being able to accurately extract
numerically convergent waveforms is crucial for any numerical-relativity
code. In the following we will show how a fully high-order scheme
improves the convergence of the inspiral and post-merger signal even when
finite-temperature EOS are used. From our simulations we extract the
$\ell=m=2$ mode of the gravitational-wave strain at a radius of $\simeq
880\, \rm km$ from the merger site for all three resolutions considered
in this work, for both the formally fourth- and second-order convergent
schemes\footnote{Ideally, since gravitational radiation is well defined
  only at future null infinity, the Cauchy evolution carried out here
  ought to be matched to a characteristic extraction evolving the
  radiation from a timelike boundary at finite radius over to future null
  infinity (see \citealt{Bishop2016} for a review). In practice, however,
  extracting at a finite radius and extrapolating to spatial infinity
  represents a very good approximation of the gravitational waveforms and
serves the scopes of this paper.}\footnote{
  The gravitational wave strain data can be downloaded from the publisher's
  website.
}

\begin{figure*}
  \centering \includegraphics[width=0.9\textwidth]{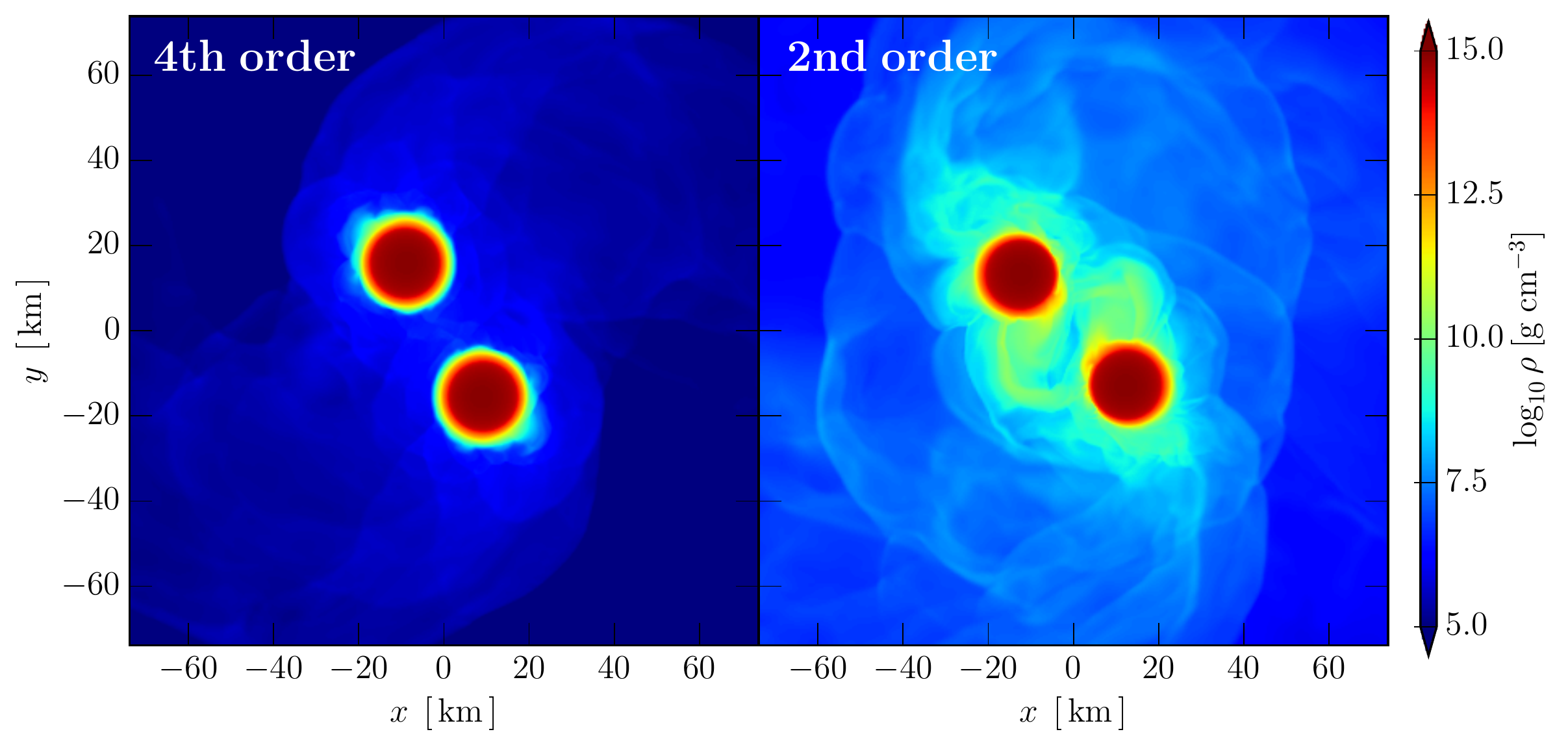}
  \caption{Rest-mass density distribution in the inspiral of an
    equal-mass neutron-star binary modelled using the finite temperature
    TNTYST EOS shortly before merger. The left panel shows the result
    using a fourth-order scheme, while the right panel shows only
    second-order accurate results. }
  \label{fig:rho}
\end{figure*}

In the top panel of Fig. \ref{fig:GW_TNTYST_strain_heff} we show the
gravitational waveforms at different resolutions (upper panel) and the
absolute error in the phase of the gravitational-wave signal, $\Delta
\phi_{22}$ (bottom panel); the latter is either measured with respect to
the highest resolution (top part) or against two neighbouring resolutions
(bottom part). In essence, we find that when using the high-order scheme
(solid lines) consistent phase evolutions are achieved even after merger.
When instead using the second-order scheme (semi-transparent lines), we
find that although convergence is obtained in the inspiral, the phase
evolution shows non-convergent behaviour after merger. On the contrary,
for the high-order scheme, consistent and convergent behaviour is also
found for the lowest resolution.

To quantify the accuracy of the waveforms the right
panel of Fig. \ref{fig:GW_TNTYST} shows the point-wise self-convergence
order for both schemes and the lower panel shows relative phase differences
$\Delta \phi_{22}^{\rm relative}$ rescaled to a given convergence order
consistent with the pointwise self convergence order in the right panel.
In the case of the fourth-order scheme (solid
line) we find that while in the inspiral a convergence order of only
$2.5$ is achieved, shortly after the merger transient, constant
third-order convergence is established. We point out that this matches
the convergence order of the employed Runge-Kutta time integration
scheme, while our spatial scheme is formally fourth-order convergent. We
attribute the reduced convergence order during the inspiral phase to a
loss of accuracy at the sharp surface of the star, which we will discuss
in the following section. Not being able to reach high-order convergence
in the inspiral gravitational-wave signal is not unprecedented and has
been studied in detail \citep{Bernuzzi2016}.

In order to provide a better comparison to previous results and also to
further assess our ability to compute highly accurate results, in
App. \ref{app:GW} we show the results obtained for one model of
\citep{Bernuzzi2016}. Considering the second-order scheme, we indeed find
second-order convergence in the inspiral, while the post-merger is no
longer convergent as discussed before.

Previous studies of post-merger simulations have established that the
frequencies of the post-merger signal obey universal relations
\citep{Bauswein2012a,Takami:2014,Takami2015,Rezzolla2016} and hence can
be used to identify the EOS \citep{Bose2017} once it will be detectable
with future detectors. In the bottom panel of
Fig. \ref{fig:GW_TNTYST_strain_heff} we report the frequency spectrum for
the gravitational-wave signals of our simulations, distinguishing again
between results for the high-order scheme (top panel) and second-order
scheme (bottom panel). The cyan solid line represents the sensitivity
curve of advanced LIGO, while different symbols mark the frequency at
merger and the frequency of the $f_2$ peak \citep{Takami:2014}. Although
all spectra agree well across all resolutions, small differences are
visible and highlight the accuracy of the high-order numerical scheme. In
particular when considering the minimum around $3\, \rm kHz$ and the
second frequency peak around $4.5\, \rm kHz$, one can see that the lowest
resolution is not converged, whereas for the high-order scheme all
resolutions yield consistent spectra across the relevant frequency range
and with the main spectral features (i.e., the $f_1, f_2$ and $f_3$
peaks) clearly visible.

Apart from being able to extract convergent gravitational waveforms it is
also important to set accurate initial conditions for the post-merger
evolution and to clarify the effect of tidal interactions on the neutron
stars, e.g., in the context of proposed pre-merger amplifications of the
magnetic field \citep{Ciolfi2017}. One aspect of the inspiral different
from the late-time evolution of the system is the presence of sharp
gradients at the surfaces of the neutron stars. This region is prone to
cause spurious mass outflows, as the hydrodynamical fluxes and the
gravitational sources do not exactly balance, resulting in unphysical
conserved states and inversion failures. As an example we show in
Fig. \ref{fig:rho} the rest-mass density distribution shortly before
merger. We find that when using a second-order accurate scheme (right
panel) that large amounts of spurious matter contaminate the domain close
to the merger and numerical inversion failures at the surface break the
initial symmetry of the equal mass system. We find that this seems to
worsen when strong magnetic fields are present in the neutron stars and
also when finite-temperature EOS are used. In contrast, when using a
fully fourth-order convergent scheme, i.e., \eqref{eqn:FD_DZ}, we find
that this spurious mass outflow is entirely absent. We also find that the
surface of the two stars show no signs of inversion failures and the
symmetry of the initial conditions is well preserved.

\subsection{Merger dynamics: Mass ejection}

\begin{figure*}
  \centering \includegraphics[width=0.95\textwidth]{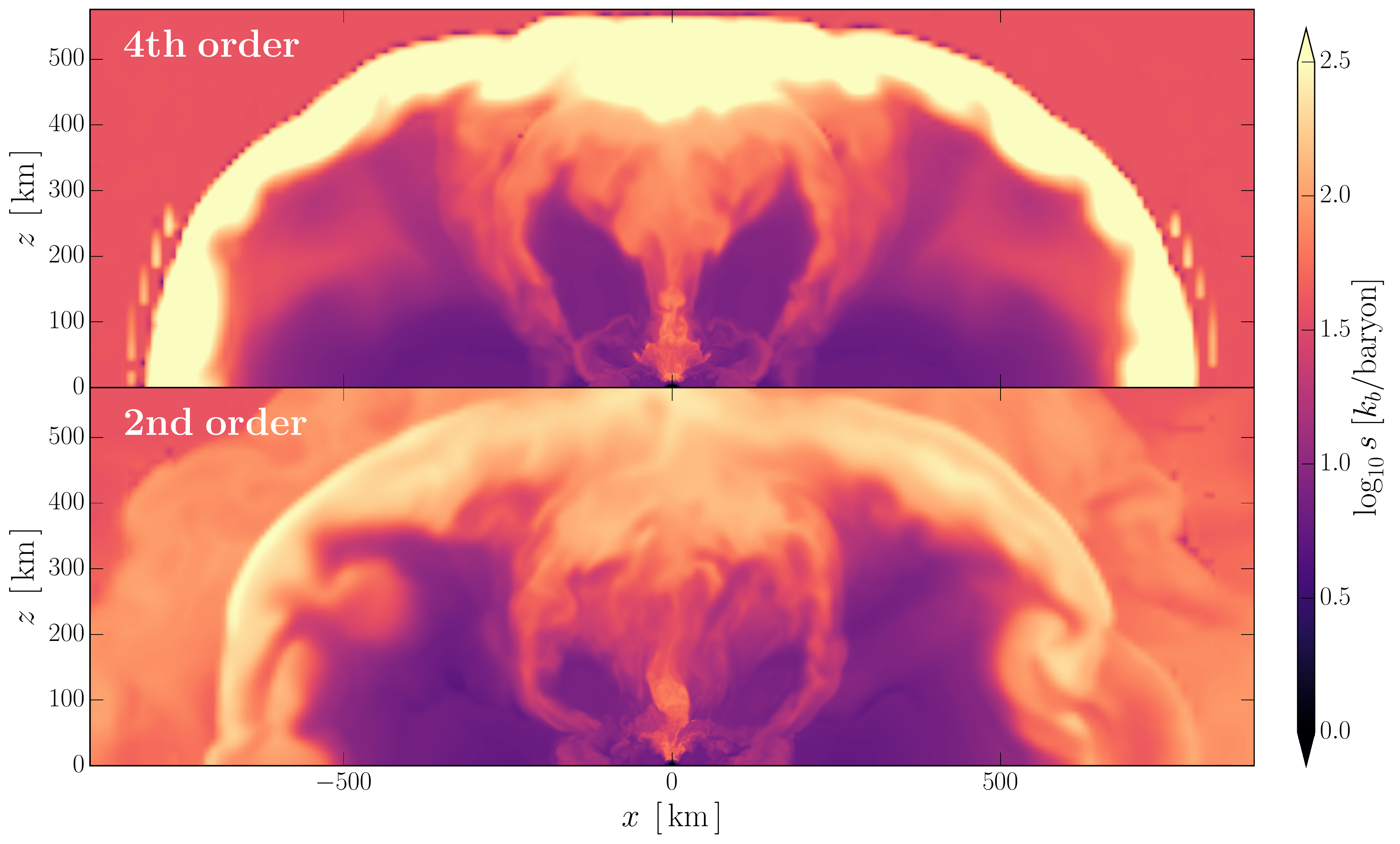}
  \caption{Entropy per baryon $s$ in the meridional plane $<10\,\rm ms$
    after merger. The top panel shows the evolution of the system for the
    fourth- order accurate scheme, the bottom panel for the second-order
    scheme. The second-order scheme introduces asymmetries and fluid
    instabilities in the flow that are entirely absent when using the
    fourth-order scheme.}
  \label{fig:s}
\end{figure*}

With the detection of the neutron-star merger event GW170817
\citep{Abbott2017_etal} and the subsequent kilonova AT2017gfo
\citep{Cowperthwaite2017,Drout2017} modelling the mass ejection from
merging binary neutron stars has become a very important part of
numerical studies. The composition of the ejected mass depends
sensitively on the precise inclusion of weak interactions in merger
simulations and considerable effort has been aimed at studying the impact
of different approaches to neutrino-transport. In line with other works
taking a more simplified approach \citep{Palenzuela2015, Lehner2016,
  Bovard2017, Radice2018a} we here explore the impact of having accurate
fluid dynamics when using finite temperature EOS in the presence of
magnetic fields.  In particular, we will focus on the dynamical part of
the mass ejection stemming from the first $20-30\, \rm ms$ after the
merger.  Although the dominant part of the mass ejection is produced on
secular time scales $>200\, \rm ms$, studying the impact of high-order
methods on long-term disk evolutions is much more involved and requires a
very careful analysis on its own \cite{Porth2019}, so that we reserve it
for a future study.

During and shortly after the merger large amounts of matter are ejected
either by tidal interactions of the two stars or by shock heating at and
after the merger. The former will lead to mass outflows mainly in the
equatorial plane, while the latter mass ejection is mainly
isotropic. Given the small sizes of the two stars and the strong shocks
involved in the process, the amount of ejected mass can quite sensitively
depend on the accuracy of the numerical scheme, the resolution and the
criterion to determine unbound material
\citep{Sekiguchi2016,Bovard2016,Bovard2017,Radice2018a}. In the
following, we will compare the effects of resolution and of using the
fourth-order scheme on the mass ejection of the equal mass system.

\begin{figure*}[h]
  \centering
  \includegraphics[width=0.9\textwidth]{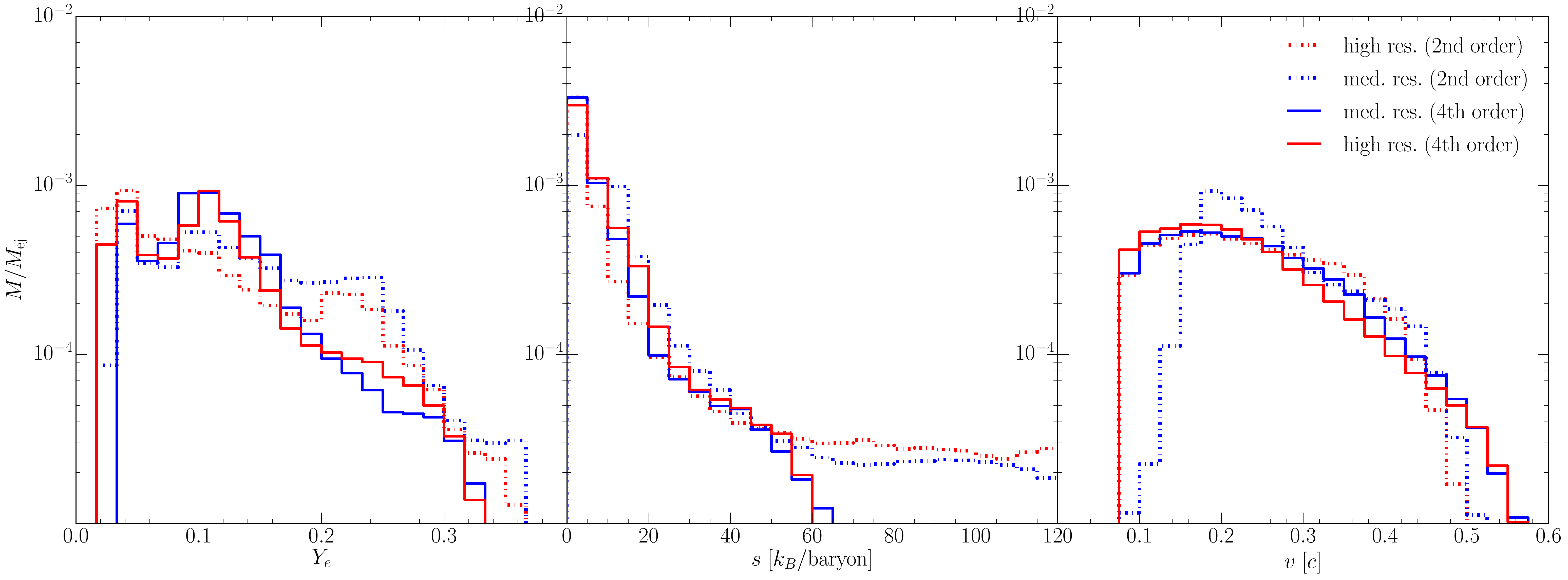}
  \caption{Distributions of the composition features of the ejected
    material. From left to right: the electron fraction $Y_e$, the
    entropy $s$ per baryon and the velocity $v$ as computed from the
    Lorentz factor of the outflow. Solid lines refer to results computed
    using the fourth-order scheme, dashed lines to results from the
    second-order scheme.}
  \label{fig:prop}
\end{figure*}

We start by looking at the shock front of dynamical mass ejection shortly
after merger, shown with its entropy values in Fig. \ref{fig:s} in the
meridional plane. It is easy to distinguish the tidally driven part of
the ejection at low entropy, from the shock heated polar driven part at
high entropy. We can also see that the ejection is preceded by a fast,
low density shock front with very high entropy values. Comparing the top
panel computed with the fourth-order scheme with the bottom panel using
the second-order scheme, we find several important differences. Firstly,
the outer shock front is Rayleigh-Taylor unstable in both cases, but the
fourth-order scheme manages to resolve it very sharply even as the matter
crosses refinement levels of decreasing numerical resolution. In the
simulation using the second-order scheme, instead we find large scale
instabilities in the fluid flow and some artificial low density outflow
preceding the shock front, which is entirely absent in the high-order
case. Secondly, starting from an equal mass binary introduces a symmetry
in the system, and apart from symmetry breaking effects like an $m=0$
instability \citep{Paschalidis2015,East2016}, the numerical scheme should
be able to preserve it over time until large scale turbulence
appears. Comparing the plumes of ejected matter in the polar direction we
find that numerical error in the second-order scheme triggers a kink-like
instability and produces asymmetries in the plume already $<10\,\rm ms$
after merger. On the contrary, the high order scheme produces a perfectly
symmetric outflow and plume, although the symmetry is never manually
imposed in the simulation.

\begin{figure}
  \centering
  \includegraphics[width=0.49\textwidth]{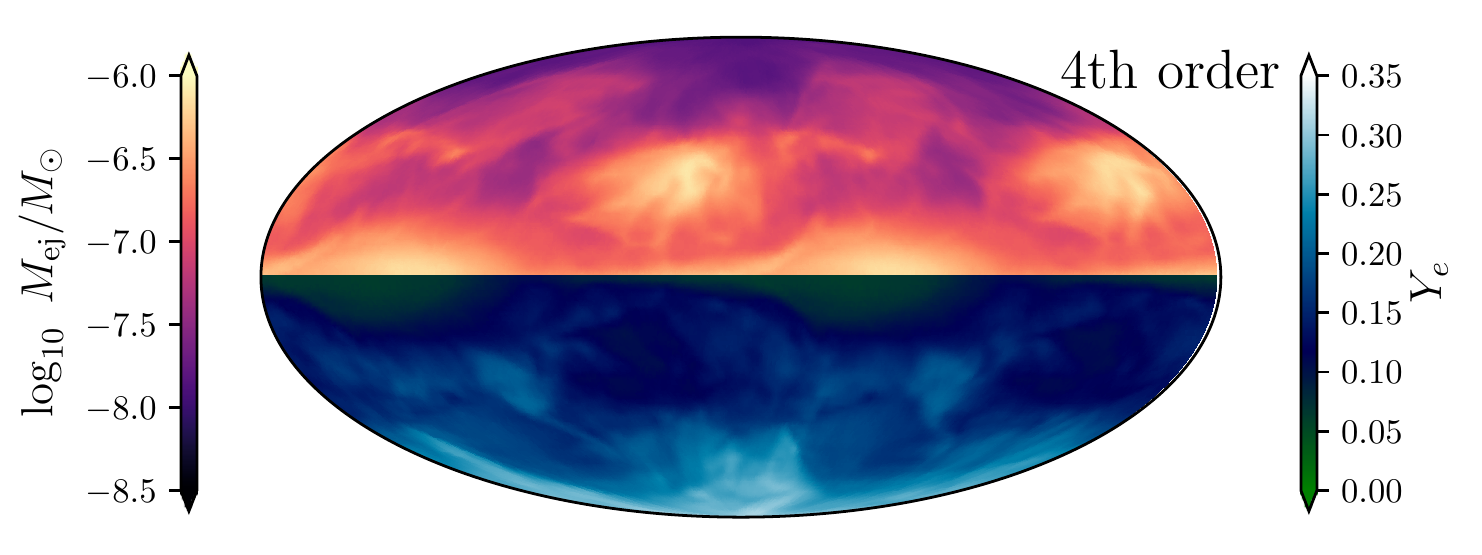}
  \includegraphics[width=0.49\textwidth]{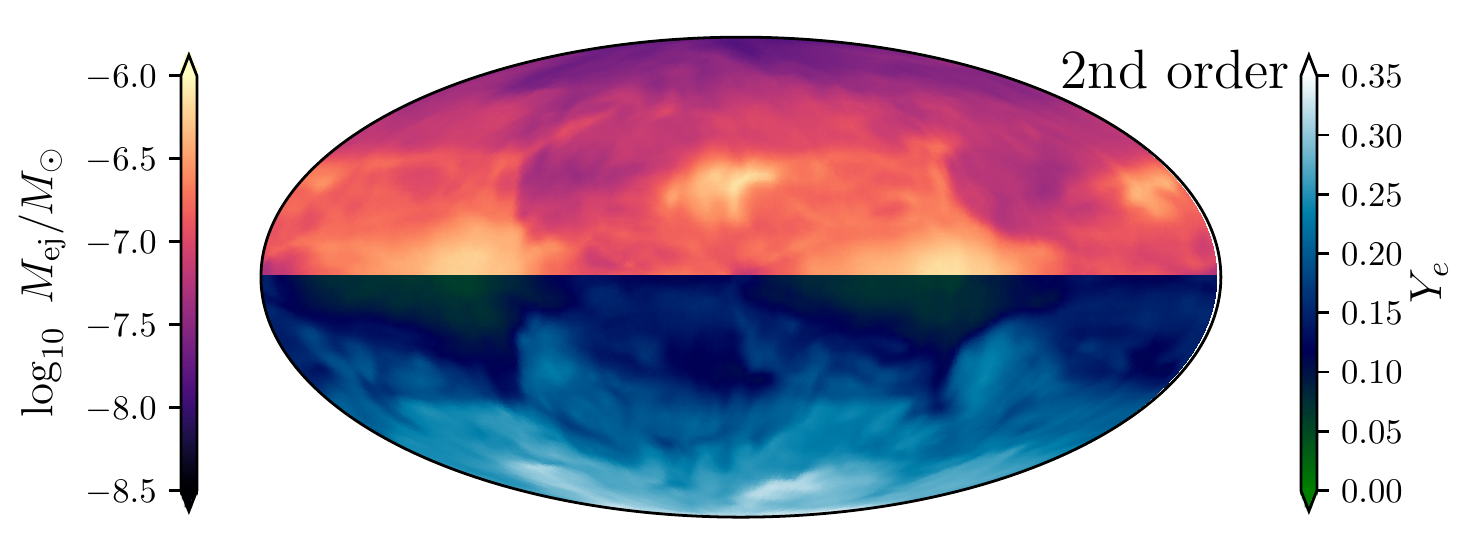}
  \caption{Time integrated ejected mass $M_{\rm ej}$ and mass weighted
    electron fraction $Y_e$ computed using the fourth-order (top) and
    second-order schemes (bottom).}
  \label{fig:Det}
\end{figure}

In order to quantify the effects of the scheme on the properties of the
mass ejection we consider the time integrated mass ejection on a sphere
placed at $\simeq 738\, \rm km$ from the origin.  We find that the
spatial distribution of the ejected matter is qualitatively the same for
both schemes, although it appears to be slightly more smeared out in the
case of the second-order scheme.  Considering the composition of the
ejected matter we find that the time and mass averaged electron fraction
$Y_e$ is significantly higher in polar regions for the second-order
scheme, reflecting its drawbacks in the modelling of polar ejecta as
anticipated from Fig. \ref{fig:s}. To better understand and quantify this
difference we consider the distributions of the ejected mass in terms of
electron fraction $Y_e$, entropy per baryon $s$ and velocity $v$. This is
shown in Fig. \ref{fig:prop}, which reports from left to right: the
electron fraction $Y_e$, the entropy $s$ per baryon and the velocity $v$
as computed from the Lorentz factor of the outflow; the solid lines refer
to results computed using the fourth-order scheme, dashed lines to
results from the second-order scheme. All simulations peak around
$Y_e\simeq 0.1$ corresponding to the tidally driven ejecta (see also
green regions in Fig. \ref{fig:Det}).  Additionally, we find that the
distribution of the electron fraction for the second-order scheme has a
second peak around $Y_e\simeq 0.25$, which is entirely absent in the case
of the high-order scheme. When comparing this feature at the two highest
resolutions we find that not only is the overall distribution of the
high-order scheme unchanged, but also that the second-order scheme even
at high resolutions overestimates the amount of proton rich
material. Further we find that the ejecta velocity are only correctly
reproduced at the highest resolution for the second-order scheme, whereas
the fourth-order scheme yields consistent distribution already at medium
resolution. In line with the discussion of Fig. \ref{fig:s} we find that
the second-order scheme produces a large tail of high entropy ejecta even
at high resolution, which again is not present in the high-order
simulations that feature a sharp cut-off around $60\, \rm k_B/baryon$.

\begin{figure*}
  \centering \includegraphics[width=0.98\textwidth]{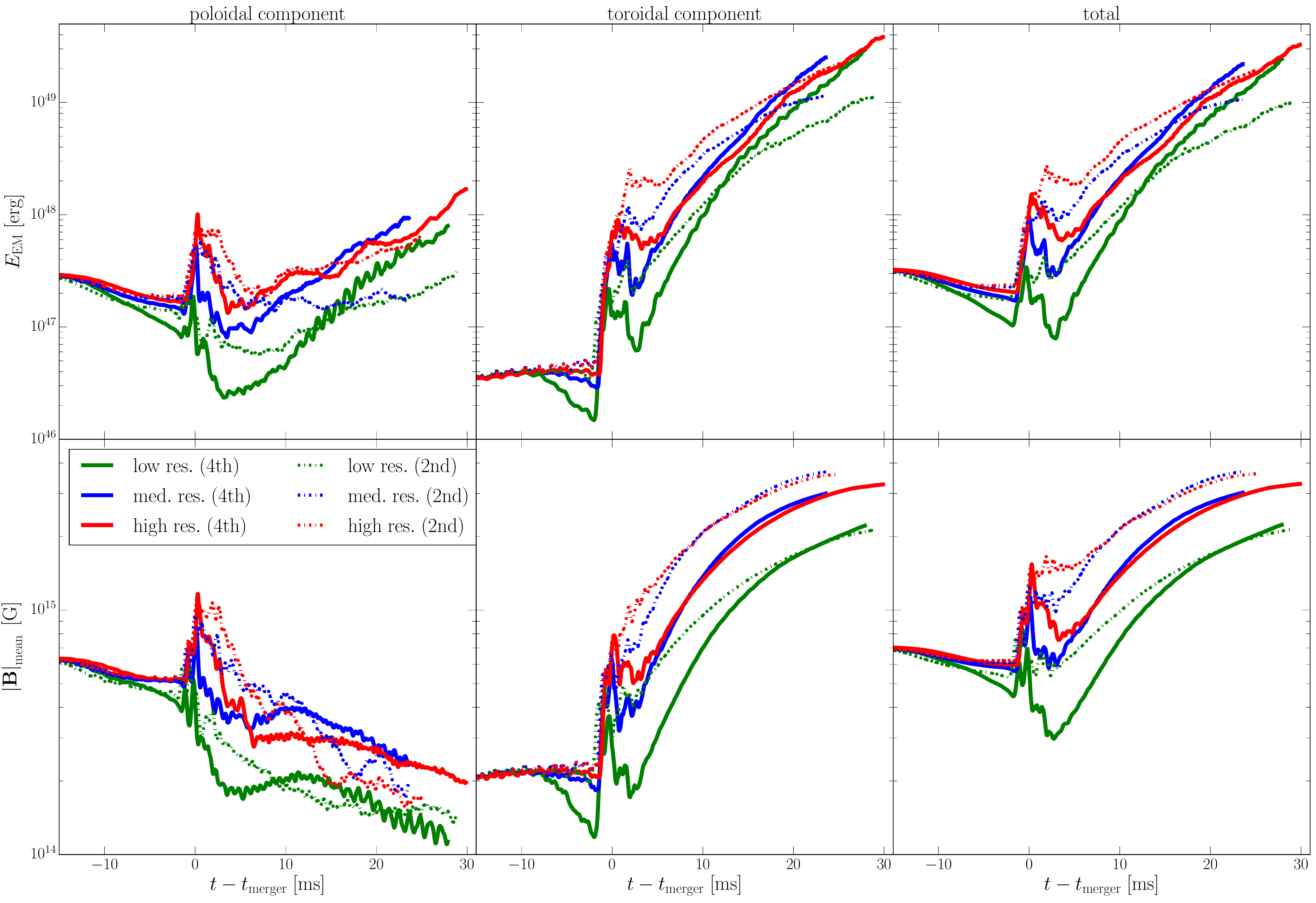}
  \caption{Magnetic field amplification in the merger remnant showing the
    poloidal and toroidal components as well as their sum. (Top row)
    Electromagnetic energy (Bottom row) Density weighted average magnetic
    field}
  \label{fig:em_energies}
\end{figure*}

\subsection{Magnetic field amplification in the merger remnant}

After the merger a differentially rotating hypermassive neutron star
(HMNS) is formed, which has a core that spins slower than its outer
layers \citep{Kastaun2014,Kastaun2016,Hanauske2016}. At the same time
also a disk is formed around the HMNS and various magnetic field effects
will cause an amplification of the magnetic field in different parts of
the merger remnant. Examples include magnetic braking \citep{Shapiro00}
which removes differential rotation and generates a toroidal field inside
the remnant, whereas the magneto-rotational instability (MRI)
\citep{Balbus1991,BalbusHawley1998} quickly leads to a magnetic field
amplification and angular momentum transport in the disk. These effects
in the early post-merger have been studied in great detail
\citep{Kiuchi2013,Endrizzi2016,Ciolfi2017} and we here focus only on the
benefits of using high-order methods to study the evolution of the
remnant.

In Fig. \ref{fig:em_energies} we report the evolution of the poloidal and
toroidal components of the electromagnetic energy $E_{\rm EM}$ and of the
density weighted magnetic field
\begin{align}
  B_{\rm mean} =\frac{\int {\rm d}V\ \rho_\ast
    \left|\mathbf{B}\right|}{\int {\rm d} V \rho_\ast}\,,
  \label{eqn:Bmean_def}
\end{align}
describing the evolution of the magnetic field inside the HMNS. Since we
start with a poloidal field confined to the interior of the two stars,
the field will simply be advected during the inspiral and we do not find
any sign of a proposed pre-merger amplification \citep{Ciolfi2017}. At
the time of merger a magnetic field amplifying Kelvin-Helmholtz
instability is expected to set in \citep{Price06,Kiuchi2015}, which is,
however, not present in any of our simulations, due to a lack of
resolution. Shortly after the merger when the HMNS is formed magnetic
braking sets in and reduces the amount of poloidal magnetic field, as can
be seen from the mean magnetic field strength in the lower left panel. At
the same time a toroidal magnetic field is generated and leads to an
amplification of the overall magnetic field strength in the remnant.
Comparing this behaviour with the exponential growth of the
electromagnetic energy reveals that at the same time an amplification,
presumably by the MRI, is active in the disk. For the different
resolutions for the fourth-order scheme we find that for each resolution
(even for the lowest one) consistent growth rates of the magnetic field
strength and energy are observed in all cases. Taking a closer look at
the poloidal energy, we can observe that an exponential growth is present
that is only correctly reproduced at the highest resolution for the
second-order scheme (red dash-dotted curve), whereas for the high-order
scheme all resolutions yield the same behaviour. This could be the result
of a minimum resolution requirement to resolve the MRI
\citep{Siegel2013}, which is expected to be much weaker when more
accurate high-order schemes are used. A comparison with the mean poloidal
field inside the remnant reveals that this growth mainly originates from
the disk, where lower resolutions are employed and hence high-order
schemes are crucial in still capturing the essential features of the
evolution of the system.

\begin{figure*}
  \centering \includegraphics[width=0.98\textwidth]{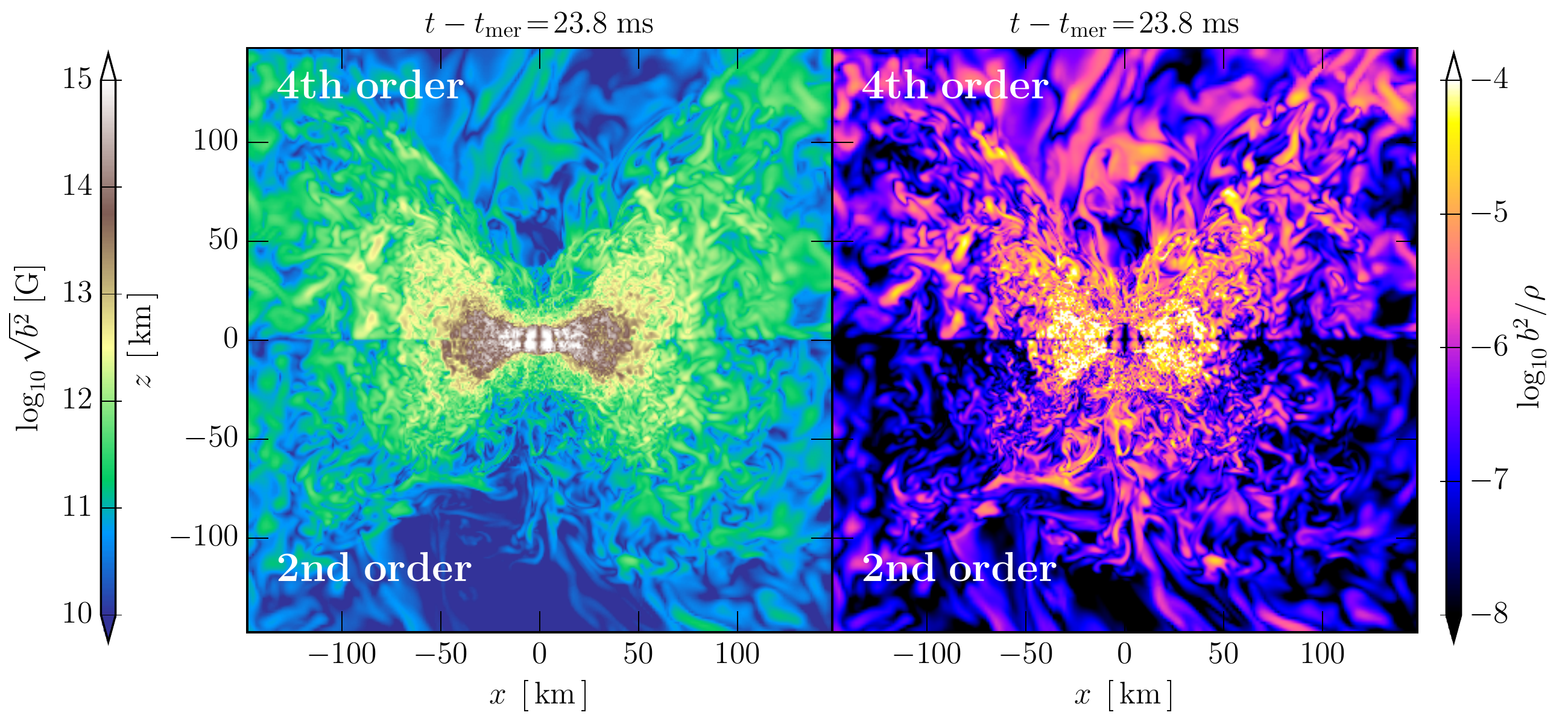}
  \caption{(Left) Magnetic field strength $\sqrt{b^2}$ in the comoving
    frame. (Right) Magnetisation parameter $b^2/\rho$. The upper part
    shows the fourth-order , the lower part the second-order
    results. Both panels refer to the high resolution case.}
  \label{fig:b2_2D}
\end{figure*}

Regarding the evolution of the toroidal component we find that inside the
remnant consistent growth rates and field strength are produced for both
fourth- and second-order scheme, but the overall growth rates of the
toroidal energy appear different for the second- and forth order case.
For the two highest resolutions the magnetic field strength amplification
inside the remnant is actually almost the same as can be seen from the
overlap of the red and blue curve in the bottom middle panel of
Fig. \ref{fig:em_energies}, different from what has been found in similar
studies using polytropic EOS and only second order accurate schemes
\citep{Endrizzi2016,Ciolfi2017}. In order to better clarify the spatial
differences in the magnetic field evolution Fig. \ref{fig:b2_2D} shows
the magnetic field strength in the comoving frame, $b^2$, (left panel)
and the magnetisation $b^2/\rho$ (right panel) for both the fourth-order
(top) and the second-order (bottom) accurate schemes. As we have already
seen from Fig. \ref{fig:em_energies} the magnetisation and field strength
inside the remnant is similar but there are large differences in the
disk. As can be clearly seen on the the right panel the high-order
simulation produces a consistent magnetisation across the domain and
specifically inside the disk on scales $>100\, \rm km$. In contrast, in
the second-order scheme the disk close to the equatorial plane is several
orders less magnetised indicating a strong lack of resolution even in the
case of our highest resolution simulation.


\section{Conclusions}

We have investigated the impact of using a forth-order accurate numerical
scheme in the simulation of merging magnetised neutron stars including
finite-temperature EOS and neutrino cooling. We have presented the
Frankfurt/IllinoisGRMHD (FIL) code, which extents the original
IllinoisGRMHD code \citep{Etienne2015} with a fully fourth-order
numerical scheme based on the ECHO approach \citep{DelZanna2007} and
implements a neutrino leakage scheme with finite temperature EOS support
combined with improved primitive inversion methods. Since the
fourth-order scheme requires the use of one additional ghost zone per
direction and the computational of the associated flux, the fourth-order
scheme is accompanied by an increased computational cost of $\sim30\%$,
as measured in our current implementation. However, these additional
costs are easily compensated by the gains in accurateness and consistency
of the solution.

Using \texttt{FIL} we have demonstrated first that we can capture the
inspiral dynamics better by significantly reducing the amount of
inversion failures at the surface of the neutron stars and second our
ability to compute accurate gravitational waveforms even when employing
realistic microphysics. More specifically, we have shown that using
\texttt{FIL} can obtain a convergence order of $\simeq 2.5$ for the phase
of the gravitational-wave signal during the inspiral and reach third
order convergence in the post-merger phase, which is most likely only
limited by the strongly stability preserving third order time
integration.

With the advent of multi-messenger astronomy of neutron-star mergers and
more detections being expected soon to follow, accurate modelling of mass
ejection has become a central part in the study of neutron-star mergers.
We have investigated the impact the fourth-order accurate numerical
scheme on the dynamical ejection of mass following the merger. We were
able to show that the outgoing shock front is much more accurately
captured, while the second-order simulations suffers from large scale
Rayleigh-Taylor instabilities and do not well preserve the initial
symmetry of the equal mass system. When considering the properties and
composition of the ejected mass, the second-order scheme produces
spurious high entropy ejecta and overestimates the amount of proton rich
material.

Finally, when considering the magnetic field evolution we found that the
fully fourth-order accurate approach allowed us to resolve poloidal and
toroidal field amplification in the merger remnant showing consistent
growth rates even at low resolution. The second-order simulation, on the
other hand, showed no poloidal field amplification and saturated early in
the toroidal field. While further investigations on longer timescales and
at higher resolutions for the second-order scheme will be necessary, we
believe that our results already indicate the importance of considering
high-order schemes for GRMHD simulations of neutron stars, especially
when considering long-term post-merger simulations.

One of the points not addressed in this study is the long-term evolution
of the system, specifically of the disk. Such systems
\citep{Siegel2017,Siegel2018,Fernandez2018} are crucial for explaining
the observed kilonova AT2017gfo as they provide large amounts of neutron
rich ejecta in the equatorial plane. Since the timescale of this ejection
is $\sim 1\, \rm s$, errors of the numerical scheme will accumulate over
time. Hence, it will be important to assess the reliability of the
current evolution scheme also for these systems, which we reserve for a
future study.

\section*{Acknowledgements}
ERM and LJP acknowledge support from HGS-HIRe. We thank Fabio Bacchini,
Oliver Porth, Hector Olivares, and Bart Ripperda for useful discussions.
Support comes in part from HGS-HIRe for FAIR; the LOEWE-Program in HIC
for FAIR; ``PHAROS'', COST Action CA16214 European Union's Horizon 2020
Research and Innovation Programme (Grant 671698) (call FETHPC-1-2014,
project ExaHyPE); the ERC Synergy Grant ``BlackHoleCam: Imaging the Event
Horizon of Black Holes'' (Grant No. 610058); The simulations were
performed on SuperMUC at LRZ in Garching, on the GOETHE-HLR cluster at
CSC in Frankfurt, and on the HazelHen cluster at HLRS in Stuttgart.

\bibliographystyle{mnras}

\bibliography{aeireferences}

 \appendix

 \section{gravitational-wave convergence: Long inspirals}
 \label{app:GW}

\begin{figure}
   \centering
   \includegraphics[width=0.42\textwidth]{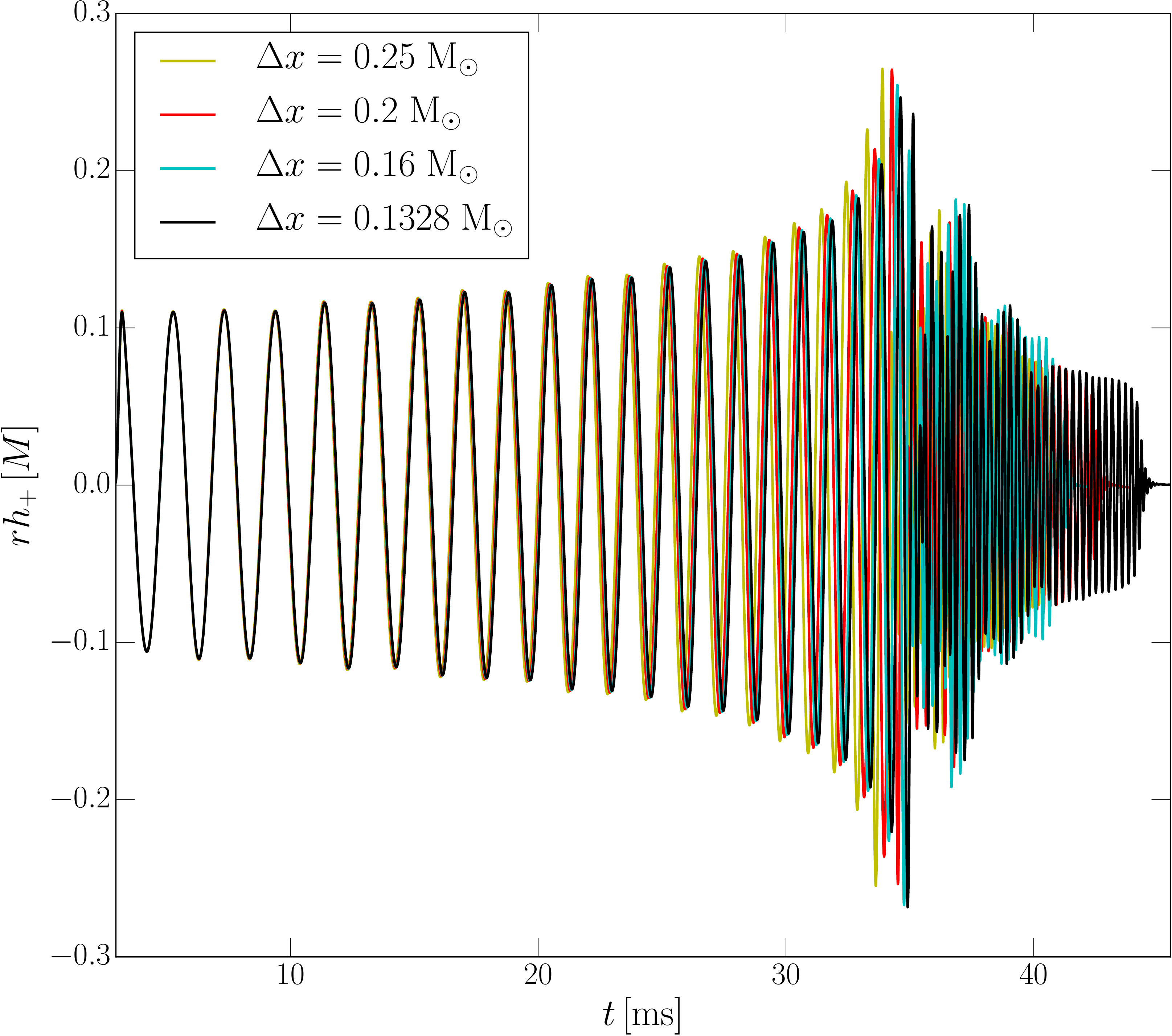}
   \hspace{7pt} \includegraphics[width=0.41\textwidth]{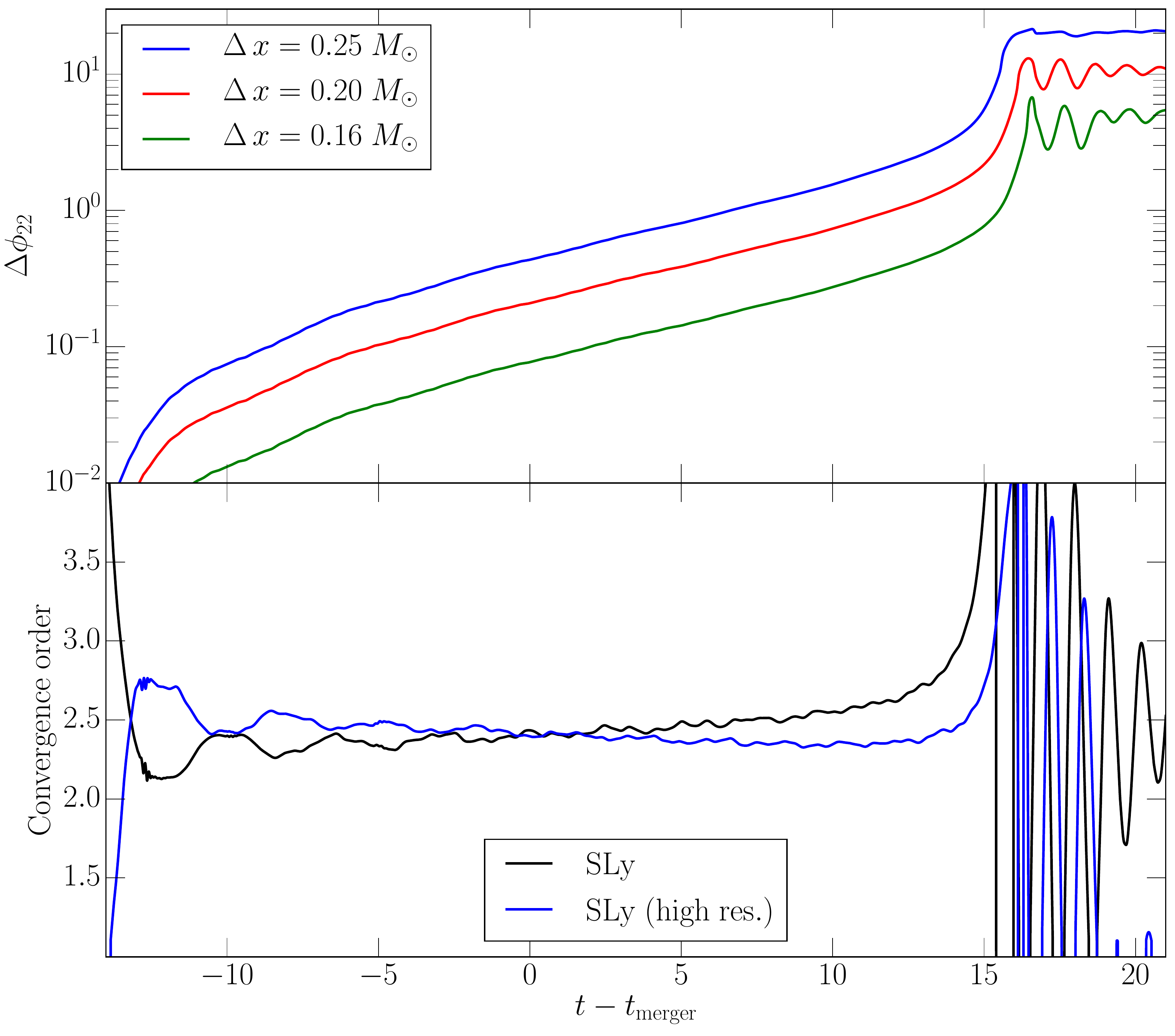}
   \caption{(Left) gravitational-wave strain $(l=m=2 {\rm mode})$
     extracted at $\simeq 880\,\rm km$ for different resolutions ranging
     from $0.1328-0.25\, M_\odot$. (Right) Relative convergence order
     computed of the gravitational-wave phase using the
     $\left(0.16,0.2,0.25\right)\, M_\odot$ and
     $\left(0.1328,0.16,0.2\right)\, M_\odot$ (HR case).}
   \label{fig:gw_comparison}
 \end{figure}

 In order to check the accuracy of the \texttt{FIL} code in extracting
 gravitational-wave signals, we study the inspiral of an equal mass
 binary with a total mass of $2.7 M_\odot$ using the SLy EOS
 \citep{Douchin01} that is initially placed in a quasi-circular orbit at
 a separation of $60\,\rm km$ \citep{Bernuzzi2016}. The simulations were
 performed on a series of nested equally spaced grids extending up to
 $\simeq 1500\,\rm km$, with four resolutions
 of\\ $\left(0.1328,0.16,0.2, 0.25\right)\, M_\odot$ on the finest
 grid. In Fig. \ref{fig:gw_comparison} we show the extracted
 gravitational-wave strain and the relative convergence order computed
 for the gravitational-wave phase at every time for two subsets of the
 resolution. We can see that the gravitational waveform is nicely
 convergent in the inspiral and difference only appear in the post-merger
 phase. Nonetheless, at all resolutions a BH is formed shortly after
 merger. We find that convergence in the inspiral is obtained even for
 the lowest resolution and that the same convergence order is obtained
 for the two subsets. Although our scheme is formally fourth-order
 accurate, we find that the convergence order as computed for two subsets
 of the resolution is only $\simeq 2.5$, similar to what has been
 observed for the same configuration in a formally fifth-order convergent
 code \citep{Bernuzzi2016}.

\bsp 
\label{lastpage}
\end{document}